\documentclass[prx,twocolumn,unsortedaddress,nofootinbib,superscriptaddress]{revtex4-2}
\usepackage{newtxtext}
\usepackage{graphicx} 
\usepackage[colorlinks=true,linkcolor=blue!60!black,citecolor=blue!60!black, urlcolor=blue!60!black]{hyperref}
\usepackage[utf8]{inputenc}
\usepackage{amsmath}
\usepackage{amssymb}
\usepackage{amsthm}
\usepackage{mathtools} 
\usepackage{physics}   
\usepackage{bm}        
\usepackage{dsfont}    
\usepackage{mathrsfs}  
\usepackage{wasysym}   
\usepackage{graphicx}
\usepackage{xcolor}
\usepackage{float}
\usepackage[normalem]{ulem}
\usepackage{soul}
\usepackage{verbatim}
\usepackage{tabularray}
\usepackage{appendix}
\usepackage{chngcntr}
\usepackage{hyperref}
\usepackage{cleveref}

\usepackage{tikz}
\renewcommand{\d}{\mathop{}\!\mathrm{d}}
\usepackage{siunitx}
\usepackage{lipsum}

\begin{document}

\title{Temporal information processing on a $4{,}500$-qubit quantum annealer}

\author{Antonio Sannia}
\email{sannia@ifisc.uib-csic.es}
\affiliation{Institute for Cross-Disciplinary Physics and Complex Systems (IFISC) UIB-CSIC,
Campus Universitat Illes Balears, 07122, Palma de Mallorca, Spain}
\affiliation{Theoretical Division, Los Alamos National Laboratory, Los Alamos, NM 87545, USA}
\affiliation{USRA Research Institute for Advanced Computer Science (RIACS), USA}
\author{Roberto Menta}
\affiliation{NEST, Scuola Normale Superiore, I-56126 Pisa, Italy}
\affiliation{Theoretical Division, Los Alamos National Laboratory, Los Alamos, NM 87545, USA}
\author{Pratik Sathe}
\affiliation{D-Wave Quantum, 3033 Beta Ave., Burnaby, BC V5G 4M9, Canada}
\affiliation{New Mexico Consortium, Los Alamos, NM 87544, USA}
\author{Dario De Santis}
\affiliation{NEST, Scuola Normale Superiore, I-56126 Pisa, Italy}
\author{Vittorio Giovannetti}
\affiliation{NEST, Scuola Normale Superiore, I-56126 Pisa, Italy}
\author{Luis Pedro Garc\'ia-Pintos}
\affiliation{Theoretical Division, Los Alamos National Laboratory, Los Alamos, NM 87545, USA}
\author{Davide Venturelli}
\affiliation{USRA Research Institute for Advanced Computer Science (RIACS), USA}
\author{Gian Luca Giorgi}
\affiliation{Institute for Cross-Disciplinary Physics and Complex Systems (IFISC) UIB-CSIC,
Campus Universitat Illes Balears, 07122, Palma de Mallorca, Spain}
\author{Roberta Zambrini}
\affiliation{Institute for Cross-Disciplinary Physics and Complex Systems (IFISC) UIB-CSIC,
Campus Universitat Illes Balears, 07122, Palma de Mallorca, Spain}
\author{Francesco Caravelli}
\affiliation{Theoretical Division, Los Alamos National Laboratory, Los Alamos, NM 87545, USA}

\begin{abstract}
Quantum machine learning could uncover statistical structure beyond the reach of classical models, but this requires quantum models large and expressive enough to be useful and cheap enough to read out. Most approaches optimize many quantum parameters and are thus limited by expensive training loops. Here we report a quantum machine-learning model implemented on a programmable superconducting quantum annealer that processes temporal data at large scale using up to 4{,}500 qubits---the largest quantum machine-learning experiment performed to date. Following the quantum reservoir computing paradigm, the untrained native many-body dynamics generated by reverse annealing is directly used to process temporal data. We prove that the interactions produced during annealing are indispensable---a non-interacting reservoir retains no memory of its input. We evaluate our model experimentally on standard memory benchmarks and demonstrate that it can successfully forecast chaotic time series. These results establish quantum annealers as a scalable platform for large-scale quantum machine learning.
\end{abstract}

\maketitle

\section*{Introduction}

Quantum processors continue to scale rapidly in size and performance, yet identifying computational tasks that make effective use of present-day hardware remains a central challenge~\cite{Bharti}.
Among the available quantum computing platforms, quantum annealers have recently demonstrated promising capabilities in areas including quantum simulation~\cite{King2025} 
and optimization~\cite{Quinton2025},  and classical criticality analysis \cite{Sathe2026ClassicalCriticality,teza}, highlighting their potential as large-scale programmable quantum devices.

Whether these processors can also serve as an effective platform for quantum machine learning, however, remains largely unresolved. Machine learning~\cite{LeCun2015} is widely viewed as one of the most promising application areas for quantum computing because quantum systems naturally generate complex, high-dimensional dynamics that may enhance information processing~\cite{Biamonte2017, Cerezo2022}. Realizing this potential on quantum annealers requires learning architectures that naturally exploit their dynamics while remaining compatible with the constraints of the hardware.

Here we experimentally demonstrate a quantum machine learning model implemented on a D-Wave quantum annealer that, for the first time, can process temporal data at large scale using up to 4{,}500 physical qubits. This work describes the largest quantum machine learning experiment to date in terms of the number of qubits employed. Our approach follows the paradigm of quantum reservoir computing~\cite{Fujii2017, Mujal2021}, in which the intrinsic quantum dynamics of the platform are used to process temporal input data, avoiding the costly optimization procedures that often limit other quantum machine-learning approaches~\cite{Larocca2025}. Reservoir computing is generally recognized as a universal framework~\cite{Grigoryeva2018}, and as such it is a powerful approach for temporal information processing.

\begin{figure*}
    \centering
    \includegraphics[width=1.0\linewidth]{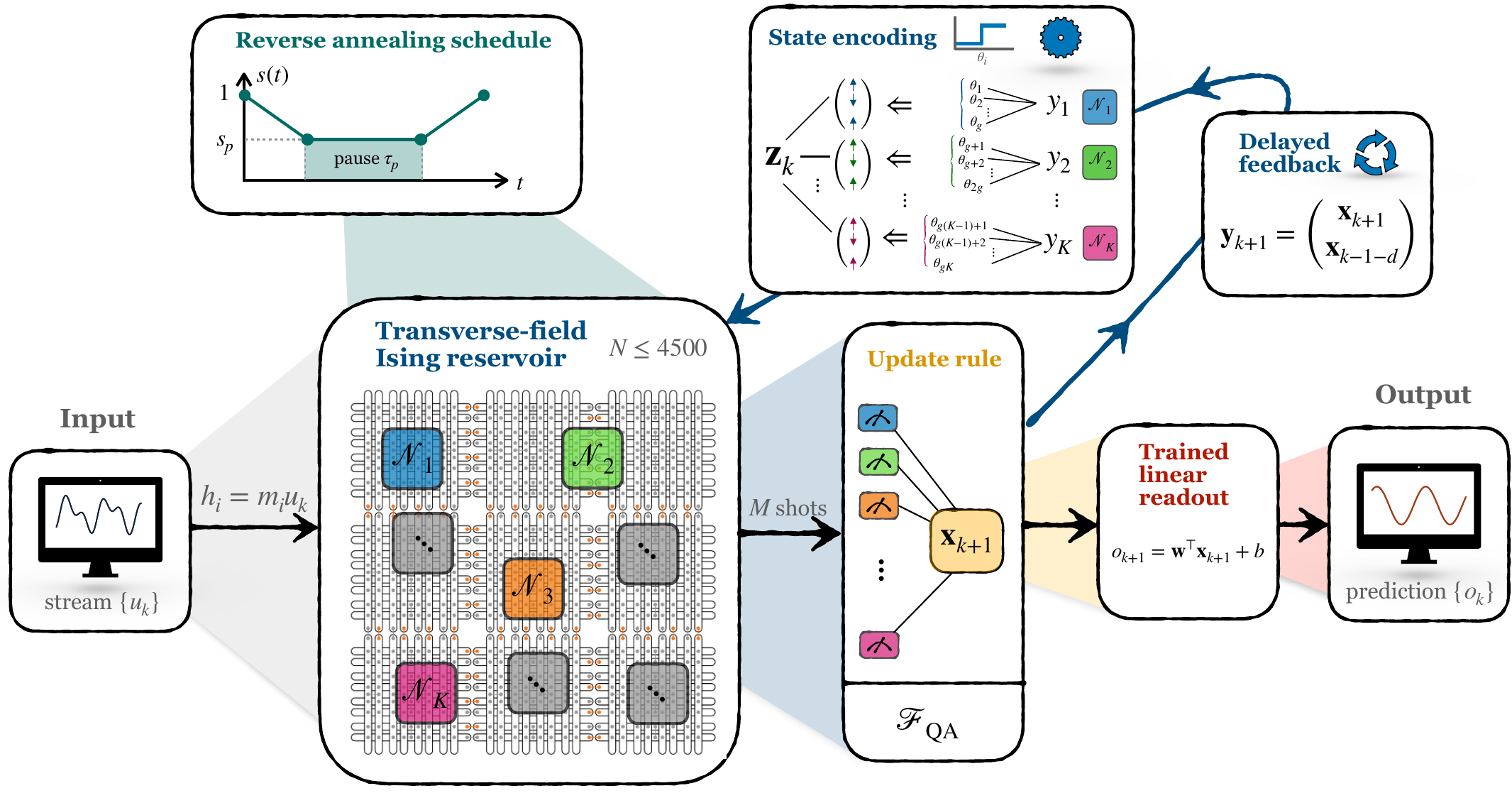}
    \caption{\textbf{Quantum reservoir computing on a quantum
    annealer.} At each time step $k$, the input $u_k$ is encoded into the local fields of the transverse-field Ising Hamiltonian via random masks, $h_i=m_i u_k$. The reservoir state is combined with components delayed by $d$ time steps and discretized using qubit-dependent random thresholds $\theta_i$ to define the initial spin configuration for reverse annealing. The final state is sampled $M$ times, and the measured spins are coarse-grained over groups of $g$ physical qubits to obtain the next reservoir state. Only the linear readout is trained, while all quantum-reservoir parameters remain fixed.}
    \label{fig:scheme}
\end{figure*}

\section*{Reservoir Computing Protocol}
The proposed quantum reservoir computing protocol consists of a vector of real numbers iteratively updated by the native many-body dynamics of the D-Wave quantum annealer~\cite{DWavePlatform}, following the scheme of quantum reservoir computers based on incoherent memory~\cite{HQRC, Kobayashi2024, Spagnolo2022, Paparelle2026}. Given $N$ physical qubits, the device realizes a transverse-field Ising Hamiltonian
\begin{equation}\label{Eq:DWaveHam}
H(s)=-\frac{A(s)}{2}\sum_{i=1}^{N}\sigma_i^x
+\frac{B(s)}{2}\Bigg(
\sum_{i=1}^{N}h_i\sigma_i^z+
\sum_{i>j}J_{ij}\sigma_i^z\sigma_j^z
\Bigg),
\end{equation}
where $\sigma_i^{x,z}$ are Pauli matrices acting on the $i$-th qubit, $s\in[0,1]$ is the annealing parameter that can be controlled as a function of time, $A(s)$ and $B(s)$ are positive hardware-specific monotonic functions such that $A(0)\gg B(0)$ and $A(1)\ll B(1)$, and $h_i$ and $J_{ij}$ are the programmable local fields and interaction strengths, respectively. 
In our application, these couplings define the reservoir connectivity.

The reservoir state at step $k$ is represented by a vector
\begin{equation}
\mathbf{x}_k=(x_{k,1},\ldots,x_{k,K}),
\end{equation}
where $K=N/g$ and $g$ is the number of physical qubits associated with each reservoir node, labeled $\mathcal{N}_\ell$, with $\ell \in \{1,\dots,K\}$. The protocol proceeds according to the following steps.

First, the reservoir state, augmented with delayed components from a previous reservoir state, is mapped onto the initial configuration of a reverse annealing protocol. This step is the encoding of the state into the machine. Specifically, we start by defining a vector $\mathbf{y}_k$ by combining the current reservoir state with a delayed one: the first half contains the components of the current reservoir state, while the second half is filled with the reservoir state from $d$ time steps earlier:
\begin{equation}
\mathbf{y}_k =
\left(
x_{k,1},\ldots,x_{k,K/2},
x_{k-d,K/2+1},\ldots,x_{k-d,K}
\right).
\end{equation}
This delayed feedback mechanism provides an explicit memory channel in addition to the memory generated by the reservoir dynamics, overcoming the memory limitations related to Markovian maps, as recently identified~\cite{Sannia2026}.

The vector $\mathbf{y}_k$ is then converted into a binary spin configuration, since the standard D-Wave quantum annealer can only be initialized in a computational-basis state and therefore requires a discrete spin assignment. The conversion is performed through the threshold rule
\begin{equation}
z_{k,i}=\Theta\!\left(y_{k,\ell(i)}-\theta_i\right),
\end{equation}
where $\ell(i)=\lfloor i/g\rfloor$ identifies the reservoir node associated with qubit $i$, $\Theta$ is the Heaviside function, and $\theta_i$ is a random threshold in $[-1,1]$. The resulting binary configuration $\mathbf{z}_k \in \mathbb{F}_2^N$ is used as the initial state of the quantum annealer.

Second, we encode the input and evolve the state.
Starting from the state just defined, a reverse-annealing protocol is executed. The system is annealed backward from $s=1$ to an intermediate point $s_p$, held for a pause time $\tau_p$, and then annealed forward again to $s=1$. Empirical evidence suggests that pausing dynamics can give rise to effective dissipative behavior~\cite{Marshall2019, Metastability}. Dissipation has also been identified as a valuable resource in quantum reservoir computing, particularly when memory is encoded directly in the quantum system~\cite{Sannia2024}. In the present hybrid setting, however, dissipation may enhance the expressivity of the quantum dynamics, playing a distinct role.

For any input value $u_k$, the local fields are programmed as
\begin{equation}
h_i(k)=m_i u_k ,
\end{equation}
where $m_i$ is a randomly chosen mask. This maps the scalar input into a global term of the quantum Hamiltonian. Generalization to multidimensional inputs is immediate via encoding in different local fields.

The resulting many-body dynamics generates a nonlinear transformation of the pair $(\mathbf{x}_{k},u_k)$.

Third, we update the state of the reservoir.
The final state is sampled $M$ times and the reservoir response is obtained from coarse-grained magnetizations,
\begin{equation}\label{Eq:Magn}
x_{k+1,\ell}
=
\frac{1}{gM}
\sum_{i\in \mathcal{N}_{\ell}}
\sum_{j=1}^{M}
\sigma_i^{z,j}(k), 
\end{equation}
where $\sigma_i^{z,j}(k)$ denotes the sampled $z$-spin value of the physical qubit $i$ in the $j$-th sample, with $\sigma_i^{z,j}(k) \in \{-1,+1\}$, obtained after the reverse-annealing evolution.

The overall recurrent update can therefore be expressed as
\begin{equation}
\mathbf{x}_{k+1}
=
\mathcal{F}_{\mathrm{QA}}
\!\left(
\mathbf{x}_{k},u_k;\mathbf{x}_{k-d}
\right),
\end{equation}
where $\mathcal{F}_{\mathrm{QA}}$ denotes the whole reservoir evolution map generated by the reverse-annealing dynamics together with the explicit delayed feedback.

The last step is the readout and training.
For a given learning task, the sequence of reservoir states is used as the feature representation of the input series, and only the output layer is trained. The corresponding outputs will be
\begin{equation}\label{Eq:outputs}
o_{k+1}=\mathbf{w}^{\top}\mathbf{x}_{k+1}+b,
\end{equation}
where $\mathbf{w}\in\mathbb{R}^{K}$ is the vector of trained readout weights and $b\in\mathbb{R}$ is a bias term. The parameters $\mathbf{w}$ and $b$ are optimized using the training data, while all reservoir parameters remain fixed.

Figure~\ref{fig:scheme} shows a schematic representation of all the evolution steps. The experimental details of the D-Wave quantum annealer configuration, together with the training and testing procedures, are provided in the Methods section.

\section*{Results}

\subsection*{Linear and non-linear memory}

\begin{figure}[t!]
    \centering  \includegraphics[width=\linewidth, keepaspectratio]{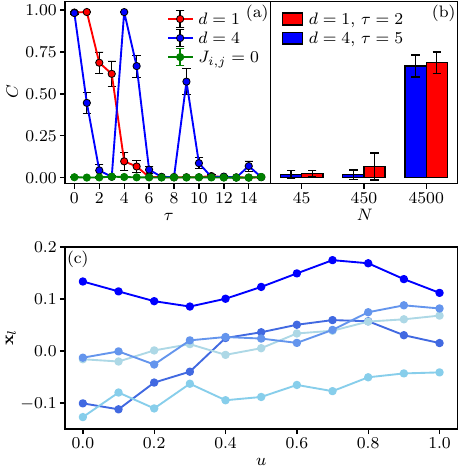}
\caption{\textbf{Reservoir response to real-valued random input streams}. (a) Linear memory capacity versus delay $\tau$ for feedback delays $d=1$ and $d=4$, for $J_{ij}\in[-0.1,0.1]$ compared with the non-interacting case $J_{ij}=0$, for $N=4{,}500$ physical qubits. 
(b) Linear memory capacity versus reservoir size $N$.  
(c) Representative reservoir-state components $\mathbf{x}_l$ after a constant input stream of value $u$, for $d=1$ and $N=4{,}500$. In all the panels, $M=10^3$ measurements and $K=45$ reservoir nodes are used. Error bars in (a)-(b) indicate the standard deviation over five independent realizations of couplings, masks, and input data.}\label{fig:LinearMemory}
\end{figure}

Having introduced the proposed architecture, we now assess its ability to retain and process temporal data. We begin by investigating its linear memory capacity~\cite{jaeger2002short}. Following standard practice in reservoir computing, the input sequence $u_k$ consists of independent random variables uniformly sampled from the interval $[0,1]$. The memory performance of the reservoir is evaluated through its ability to reconstruct past inputs. Specifically, for a given delay $\tau$, the target output at time step $k$ is defined as: $\hat{o}_k = u_{k-\tau}$, where $\tau$ determines how far into the past the reservoir must recall the input. The quality of the reconstruction is quantified by the capacity ($C$)~\cite{Dambre2012}, which coincides with the squared Pearson correlation coefficient between the predicted output and target.

Figure~\ref{fig:LinearMemory}(a) presents the linear memory capacity obtained using $4{,}500$ physical qubits to implement the reverse-annealing protocol on the 
\texttt{Advantage2\_system1} D-Wave quantum annealer. We investigated two values of the feedback delay appearing in the reservoir update rule, namely $d=1$ and $d=4$. A particularly noteworthy result is that, for $d=4$, the memory capacity exhibits clear revivals as a function of $\tau$. Such behavior can be interpreted as a manifestation of non-Markovian dynamics. Indeed, it is established that Markovian evolutions tend to erase information exponentially fast, leading to an exponential decay of the memory capacity as the delay $\tau$ increases~\cite{Sannia2026}. By contrast, the proposed architecture can overcome this limitation through the feedback mechanism, enabling the re-emergence of previously stored information and consequently generating memory revivals. 

It is also important to determine whether the reverse annealing evolution plays a meaningful role in the observed memory performance. To this end, we recomputed the linear memory capacity after setting all couplings in the Hamiltonian governing the reverse-annealing dynamics, Eq.~\eqref{Eq:DWaveHam}, to zero: $J_{ij}=0$.

Under these conditions, the qubits evolve independently, with no correlations generated among them. We find that the resulting reservoir exhibits essentially no memory of the input signal, yielding a null memory capacity and becoming ineffective for machine-learning tasks. In the Supplementary Material, we provide a proof of this behavior, showing that the correlations generated during the annealing dynamics are essential for the reservoir to retain and process temporal information. In addition, spin-vector Monte Carlo simulations~\cite{ShinSVMC}, also presented in the Supplementary Material, fail to reproduce the observed performance, suggesting that the quantum nature of these interactions may be essential.

Another important aspect concerns the role of the system dimensionality. We therefore investigated whether the large number of qubits employed is necessary to achieve the observed performance. In Figure~\ref{fig:LinearMemory}(b), we analyze the reservoir's ability to exhibit non-Markovian memory revivals by evaluating the capacity at a fixed delay (i.e., $\tau= 2 (5)$ for delay $d=1(4)$) while varying the reservoir size. More specifically, we vary the number of physical qubits over different orders of magnitude while keeping both the output layer and the number of measurements (i.e., shot noise) fixed. The results reveal that significant memory capacity at this delay emerges only when the full set of $4{,}500$ physical qubits is employed. Smaller reservoirs fail to retain correlations with the input at this timescale. Notice that since $K$ is fixed, larger $N$ also increases the number of qubits $g = N/K$ averaged per output, which lowers the shot-noise floor and may contribute to this trend alongside the dynamical role of reservoir size. This finding highlights the fundamental role played by the high-dimensional quantum reservoir in sustaining long-range temporal correlations and generating the observed memory revivals. In the Supplementary Material, we present ideal simulations of a small-scale version of the system that further confirm that this high number of qubits is necessary for the reservoir to function effectively 

Moreover, in Figure~\ref{fig:LinearMemory}(c), we show the reservoir response to a constant input stream, experimentally demonstrating that the reservoir maps the inputs into a nonlinear feature space where they become separable.

\begin{figure}[t!]
    \centering  \includegraphics[width=\linewidth, keepaspectratio]{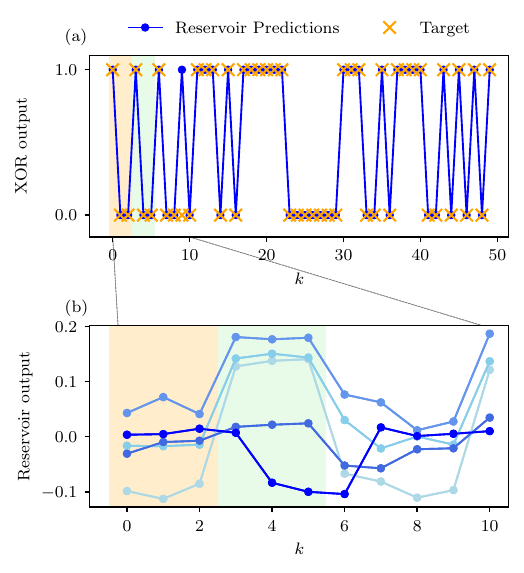}
\caption{\textbf{XOR task}. (a) Target values and reservoir predictions over discrete time steps during the test phase. 
(b) Representative components of the reservoir response over time for a smaller test segment. 
The reservoir uses feedback delay $d=1$; magnetizations are estimated from $M=10^3$ measurements to compute $K=45$ outputs.}\label{fig:XOR}
\end{figure}

To quantify the nonlinear information-processing capability of the proposed architecture, we employ the XOR task. In this setting, the input sequence consists of binary values randomly sampled from the set $\{0,1\}$. The objective is to compute the parity of the two most recent inputs injected into the reservoir. Accordingly, the target output is defined as: $\hat{o}_k = \left( u_k + u_{k-1} \right) \bmod 2$.

The XOR task is a standard benchmark for assessing nonlinear computational capabilities, as it cannot be solved through a purely linear transformation of the input data. To evaluate the reservoir performance, we employ the classification accuracy, defined as the fraction of correctly predicted outputs, over the data points in the test phase. This metric was evaluated over five independent realizations with different input sequences, Hamiltonian parameters, and threshold configurations. This yielded a value of $0.989 \pm 0.008$, where the uncertainty denotes one standard deviation across the realizations. This demonstrates the robustness of the proposed architecture and its ability to reliably capture the nonlinear relationships encoded in the input stream.

Figure~\ref{fig:XOR}(a) shows the reservoir predictions together with the target time series during the test phase for one representative realization. We observe an almost perfect overlap between the predicted and target outputs, in agreement with the high accuracy reported above. Moreover, in Figure~\ref{fig:XOR}(b), we show how a representative set of reservoir components varies over time. Two distinct regions, corresponding to the same target values, are highlighted. This shows how the performed optimization enables the reservoir to give the same correct predictions, even when its states differ.

\subsection*{Chaotic time-series forecasting}

\begin{figure*}[t!]
    \centering  \includegraphics[width=\linewidth, keepaspectratio]{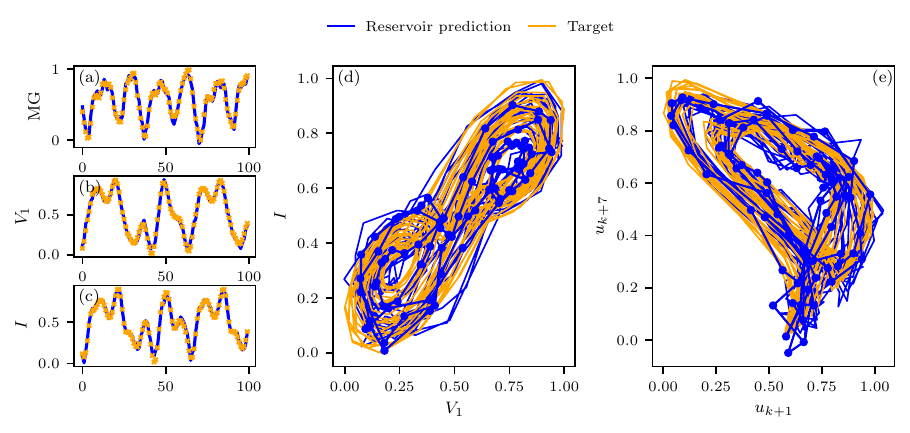}
\caption{\textbf{Chaotic time-series forecasting}. Reservoir predictions and target values for a representative 100-point segment of the test set: (a) Mackey--Glass series, and the (b) $V_1$ and (c) $I$ components of the double-scroll system. Chaotic attractors reconstructed from the Double-Scroll (d) and Mackey--Glass (e) series. The reservoir hyperparameters are $d=1$, $M=10^3$, and $K=45$. For the attractor reconstruction shown in panel (d), the output layer was built by combining the outputs of the five reservoir realizations, all driven by the same input stream. All the other panels refer to one representative reservoir realization. }\label{fig:CTS}
\end{figure*}

Having assessed the reservoir’s capacity to store and process random inputs, we now turn to the more challenging task of chaotic time-series forecasting. Specifically, we evaluate the reservoir’s ability to predict the Mackey–Glass~\cite{Mackey1977} and double-scroll~\cite{DS} time series, two well-known benchmark tests widely studied in the literature. Further details on the generation and preprocessing of these time series are provided in the Supplementary Material.

The Mackey–Glass series is a univariate signal. The reservoir is driven by its values and trained to perform a one-step-ahead prediction of the sequence, namely $\hat{o}_k = u_{k+1}$. Figure~\ref{fig:CTS}(a) shows the reservoir output and the corresponding target values for a representative segment of $100$ points from the test set. We also trained the reservoir to perform longer-horizon forecasting, with $\hat{o}_k = u_{k+7}$. In Figure~\ref{fig:CTS}(e), we plot the reconstructed one-step- and seven-step-ahead series against each other, which reveals a chaotic attractor, showing that the reservoir can successfully capture the chaotic behavior of the underlying dynamics.

In the double-scroll case, the analyzed time series consists of two components, $V_1$ and $I$. Here, too, the reservoir is trained to perform a one-step-ahead prediction task. The results for a representative segment of $100$ points from the test set are shown in Figure~\ref{fig:CTS}(b)--(c). Figure~\ref{fig:CTS}(d) shows that the two reconstructed components in the phase plane once again capture the underlying chaotic attractor of the series. To the best of our knowledge, these experiments provide the first successful reconstruction of a chaotic attractor from a quantum computer.

Finally, in the Supplementary Material, we demonstrate that the predictive performance of the proposed reservoir remains robust even for longer forecasting horizons, quantifying it with the normalized mean square error. Moreover, by comparing it with a linear autoregressive model provided with the same delayed feedback, we show that the reservoir exhibits genuine nonlinear computational capabilities.

\section*{Discussion}
We have demonstrated a quantum machine-learning model that processes temporal
data on a programmable superconducting quantum annealer, using
up to $N=4{,}500$ physical qubits---the largest quantum
machine-learning experiment reported to date by physical-qubit count. Within the quantum
reservoir-computing paradigm, the native many-body dynamics generated by reverse
annealing of a transverse-field Ising system map a stream of temporal data into a
high-dimensional feature space. Only a single linear layer is trained, while all reservoir parameters remain fixed. This design avoids the costly optimization loops and barren-plateau challenges that often hinder variational quantum machine learning~\cite{Larocca2025}. Owing to the presence of delayed feedback, the reservoir exhibits non-Markovian memory, characterized by revivals in memory capacity rather than simple exponential forgetting~\cite{Sannia2026}. Importantly, we find the many-body interactions generated during the reverse annealing are indispensable: a non-interacting reservoir ($J_{ij}=0$) retains no memory of its
input, a behaviour we trace analytically to the breakdown of the echo-state
property~\cite{Yildiz2012}. On standard benchmarks, the device solves
the nonlinear XOR task, and forecasts the well-known Mackey-Glass and double-scroll chaotic time series, outperforming a linear autoregressive model supplied with
the same delayed feedback~\cite{box1994time}.

Our findings raise the question of how the map realized by our device relates to classical computation.
As is widely recognized in the literature, it remains a major challenge to establish the classical simulability of quantum machine-learning models~\cite{Cerezo2025}. We observe that the transverse-field Ising Hamiltonian implemented by the quantum annealer is stoquastic and free of the conventional sign problem, making its equilibrium and ground-state properties amenable to sign-problem-free classical Monte Carlo methods, although stoquasticity alone does not guarantee efficient classical simulation~\cite{Bravyi2008Stoquastic}. The pertinent question is therefore not about these static quantities but about the finite-time, out-of-equilibrium dynamics that generate the reservoir map. The transformation $\mathcal{F}_{\mathrm{QA}}$ arises from a reverse-annealing quench of a $4500$-qubit system on the native, nonplanar, high-degree hardware graph. These kinds of unitary evolutions reach dynamical and critical regimes that strain state-of-the-art classical methods, including a demonstrated scaling advantage over path-integral Monte Carlo in the simulation of geometrically frustrated magnets~\cite{King2021}, coherent quantum-critical dynamics in a $5{,}000$-qubit programmable spin glass~\cite{King2023}, the reverse-annealing study of topological Kosterlitz--Thouless phenomena at the $1{,}800$-qubit
scale~\cite{King2018}, and a recent beyond-classical demonstration in programmable quantum simulation~\cite{King2025}. Reinforcing this picture, the use of a quantum annealer as a ground truth has been shown to invalidate the scaling extrapolations of state-of-the-art classical tensor-network simulations of quantum dynamics; this indicates that one should not assume that such simulations remain accurate at scale~\cite{King2025eval}. Moreover, recent theoretical results show that reverse-annealing dynamics can simulate arbitrary quantum circuits with polynomial resource overhead~\cite{werner2026}. 
These results motivate a direct investigation of the classical resources required to reproduce the present reservoir map, but they do not establish its classical intractability. Determining whether, and at what accuracy and computational cost, state-of-the-art classical dynamical simulations can reproduce the experimentally accessible reservoir map is an important direction for future work.

Beyond these questions of simulability, our results chart a practical route toward
machine learning on large-scale quantum hardware. We note that untrained quantum platforms, in which only a linear readout layer is trained, have similarly been shown to solve machine-learning problems, although at scales one order of magnitude smaller than those achieved in this work, as measured by the number of qubits or bosonic modes involved~\cite{QuEra,Cimini2026,IBM}. However, none of these studies demonstrated the ability to implement a genuine recurrent evolution, which is essential for processing time-dependent data. Consequently, these approaches are more appropriately classified as quantum extreme learning machines rather than reservoir computers~\cite{Mujal2021}.

Importantly, the experiment presented here opens the door to a different way of thinking about quantum machine learning. In the early development of the field, promising results were often obtained from small-scale numerical simulations, yet many of these results have not been shown to persist as system size increases~\cite{bowles2024betterclassical}. In the present case, we observe the opposite behavior: the proposed models become effective only at sufficiently large scales and under the experimental conditions described here, whereas idealized small-scale simulations yield systems that are unable to process time-dependent data effectively.

This model, which was identified heuristically, also appears to evade limitations associated with exponential concentration phenomena, whereby its outputs would otherwise collapse toward a fixed value at an exponential rate as the system size increases, progressively erasing input-dependent information. It is typically assumed that the unitary evolution of the quantum system is sampled from, or closely approximates, a 2-design distribution~\cite{xiong2025,sannia2025exponential}. This assumption is clearly not satisfied by the structured dynamics considered in this work. Consequently, the effectiveness of the proposed strategy cannot be inferred from those general concentration arguments and must instead be established directly through experiment.

This heuristic and experiment-driven approach is consistent with the historical development of classical deep learning, where empirical performance has often preceded a complete analytical understanding. In such settings, direct numerical and experimental evidence has provided insights that could not be obtained from complexity arguments based on idealized or insufficiently representative assumptions~\cite{LeCun2015}.

\section*{Methods}

All experiments were performed on the \texttt{Advantage2\_system1} D-Wave quantum annealer \cite{DWaveAdvantage2_2025}. The reservoir was implemented using up to $N=4{,}500$ physical qubits selected from the active qubits available on the device. The physical qubits were grouped according to the native ordering used by D-Wave to label the active qubits. After sorting the active qubit indices, consecutive blocks of $g$ qubits were assigned to the same reservoir component. Native hardware connectivity was employed throughout. The Ising couplings in Eq.~\eqref{Eq:DWaveHam} were randomly generated at the beginning of each realization and subsequently kept fixed during the entire computation. Specifically, the coupling strengths were independently sampled from a uniform distribution in the interval $[-0.1,0.1]$ for all active couplers connecting the selected qubits. Similarly, the random masks $m_i$, used for input encoding, and the thresholds $\theta_i$, employed for state discretization, were sampled respectively from the intervals $[-0.5, 0.5]$ and $[-1,1]$ and kept fixed within each realization. A new set of couplings, masks, and thresholds was generated for every independent realization. 

Reverse annealing was implemented using the schedule, with consecutive points connected by linear interpolation,

\begin{equation}
(0~\unit{\micro\second},1)
\rightarrow
(0.5~\unit{\micro\second},0.5)
\rightarrow
(1.5~\unit{\micro\second},0.5)
\rightarrow
(2~\unit{\micro\second},1).
\end{equation}

The first coordinate denotes the elapsed time, while the second specifies the annealing parameter $s$. Starting from the classical regime ($s=1$), the system is annealed backward to $s=0.5$, where quantum fluctuations are stronger due to the increased transverse-field contribution. The system is then held at this point for $1~\unit{\micro\second}$ before being annealed forward again to $s=1$ and measured. This reverse-annealing protocol allows the reservoir dynamics to explore the local energy landscape around the initialized state while preserving information about the previous reservoir configuration.

A readout thermalization time of $100~\unit{\micro\second}$ was employed after each reverse-annealing cycle. To mitigate slow temporal variations of the device parameters, flux-drift compensation was enabled throughout all experiments using the native D-Wave calibration protocol.

For all learning tasks, the readout layer of Eq.~\eqref{Eq:outputs} was trained using 500 data points. The first 10 points were discarded as a washout period to reduce the influence of the initial reservoir state and ensure operation in the echo-state regime. The readout weights were obtained through linear regression using the Moore--Penrose pseudoinverse~\cite{Penrose1955}. Performance was subsequently evaluated on an independent test set consisting of 500 additional data points. For the XOR task, during the test phase, since the target values are binary, reservoir outputs greater than or equal to 0.5 are classified as 1 while outputs below 0.5 are classified as 0.

\bibliographystyle{naturemag}
\bibliography{biblio}

\subsection*{ACKNOWLEDGMENTS}
The authors thank Cristiano Nisoli for insightful discussions. A.S. and R.M. acknowledge the New Mexico Consortium for access to the D-Wave machine. Generative AI tools were used solely to improve the English-language presentation of the manuscript and to assist in checking the numerical code. All scientific content, analyses, interpretations, and conclusions were developed and verified by the authors, who take full responsibility for the manuscript.
\subsection*{Funding}
A.S., G.L.G. and R.Z. acknowledge funding from the Spanish State Research Agency through the COQUSY project PID2022-140506NB-C21 and -C22 and the María de Maeztu project CEX2021-001164-M, and CoQuaDis project PCI2024-153446, funded by MICIU/AEI/10.13039/501100011033 and by ERDF, EU. A.S., G.L.G. and R.Z. also acknowledge MINECO through the QUANTUM SPAIN project. A.S. received the support of a fellowship from the ”la Caixa” Foundation (ID 100010434). The fellowship code is LCF/BQ/DI23/11990081. A.S. and R.M. acknowledge support from the U.S. Department of Energy (DOE) through a quantum computing program sponsored by the Los Alamos National Laboratory (LANL) Information Science $\&$ Technology Institute.
A.S. has also been supported by the USRA Feynman Quantum Academy internship program. L.P.G.P. acknowledges support from the U.S. Department of Energy, Office of Science, Basic Energy Sciences program (award No. DE-
SCL0000157). D.V. acknowledges support from DOE DE-SC0026126. 
\subsection*{Author Contributions}

A.S. conceived the work. A.S., R.M., R.Z. and F.C. firstly discussed about the possibility of using D-Wave annealers as quantum reservoirs. A.S. proposed the reservoir architecture and performed all experiments on the quantum computer, as well as the numerical simulations and data analysis. A.S. and F.C. carried out the analytical calculations. A.S. and R.M. wrote the first version of the manuscript. G.L.G., R.Z., and F.C. supervised the work. All authors contributed to the discussion and interpretation of the results and to the final preparation of the manuscript.
\subsection*{Competing Interests}
P.S. is an employee of D-Wave Quantum Inc. and declares competing interests on that basis.
While F.C. is now an employee at Planckian, this work was initiated while he was a LANL employee, where he still resides as a Guest Professor. 

\subsection*{Code availability:} 
A tutorial notebook for reproducing the experiments is available through a \href{https://doi.org/10.5281/zenodo.22127000}{Zenodo repository}.

\clearpage
\setcounter{section}{0}
\setcounter{equation}{0}
\def\theequation{S\arabic{equation}}
\setcounter{figure}{0}
\onecolumngrid
\begin{center}
\vspace*{\baselineskip}
{\Large\textbf{Supplementary  Material}}
\end{center}
\def\thefigure{S\arabic{figure}}

In the Supplementary Material, we provide additional proofs and results to support the main text.

\section{Microscopic and open-system derivation of the reservoir map}
\label{sec:micro}

In the main text, the recurrent update
$\mathbf{x}_{k+1}=\mathcal{F}_{\mathrm{QA}}(\mathbf{x}_k,u_k;\mathbf{x}_{k-d})$
is introduced operationally as the map generated by one reverse--annealing cycle followed by a coarse--grained readout. The purpose of this section is to unpack the presented model and identify the microscopic object that is iterated by the reservoir. We shall describe the same cycle at three levels of resolution. First, we treat the reverse anneal as a coherent Hamiltonian process and derive the associated hopping amplitudes on the Hamming graph by a Suzuki--Trotter expansion. Second, we replace the unitary propagator by a general completely positive trace-preserving map generated by an open-system Liouvillian. Third, in the regime of rapid dephasing, we eliminate the coherences and obtain a Pauli master equation for the populations of classical spin configurations.

This hierarchy is useful because the experiment does not require the device to remain coherent throughout the entire reverse--annealing schedule. What is required is a reproducible, input-dependent, many-body map from the prepared configuration to the measured magnetizations. Coherent tunneling, incoherent tunneling, thermally assisted relaxation, and mixed coherent--dissipative dynamics may all realize such a map. The common ingredient is that the transition amplitudes or transition rates depend nontrivially on the interacting Ising energy landscape. Consequently, the analytical results presented here demonstrate that, below any limit, the reservoir response remains a nonlinear high-dimensional input-dependent mapping, as required for a reservoir computing model.

\subsection{Reservoir computing, delayed feedback, and the one-cycle map}

Reservoir computing approximates a functional relation between an input stream and a target stream by driving a high-dimensional dynamical system and training only a simple readout~\cite{jaeger2002short}. Removing for the moment non-Markovian effects, in continuous time the reservoir evolution is governed by
\begin{equation}
 \dot{\mathbf{x}}(t)=F\bigl(\mathbf{x}(t),u(t)\bigr),
\end{equation}
where $\mathbf{x}(t)$ is the reservoir state and $u(t)$ is the external input. The reservoir need not solve the target task directly. Instead, it transforms the input history into a collection of nonlinear features, after which the target is approximated by a linear combination of them:
\begin{equation}
 o(t)=\mathbf{w}^{\top}\mathbf{x}(t)+b.
\end{equation}
Here, the reservoir output, $o(t)$, is obtained after training only the linear readout vector, $\mathbf{w}\in\mathbb{R}^{K}$, and the bias term $b\in\mathbb{R}$. The internal reservoir parameters remain fixed.

For a useful reservoir, two complementary requirements are normally imposed. The first is the echo-state property: two trajectories driven by the same input stream but initialized differently must approach one another~\cite{Yildiz2012}. Under continuity conditions typically satisfied by physical systems, the echo-state property is equivalent to the fading-memory property~\cite{Gonon2021}, whereby the reservoir state is determined primarily by the recent input history, while the influence of more distant inputs progressively vanishes.

The second is state separation: distinct recent input histories should remain distinguishable in the reservoir state. Excessive contraction in the evolution destroys state separation, whereas insufficient contraction retains an unwanted dependence on the arbitrary initial condition. The useful operating regime lies between these limits.

In the present discrete-time architecture, one reverse--annealing cycle plays the role of one application of the flow generated by $F$. Before cycle $k$, the current reservoir state and a delayed state are combined into the feedback vector $\mathbf{y}_k$, whose components are in one-to-one correspondence with the hardware physical qubits. The threshold rule
\begin{equation}
 z_{k,i}=\Theta\!\left(y_{k,\ell(i)}-\theta_i\right)
\label{Eq:ThresholdAgain}
\end{equation}
produces a binary initialization. In spin variables ($s_i=2z_i-1\in\{-1,+1\}$), this defines the computational-basis state
\begin{equation}
 \rho_{k,0}=\ket{\mathbf{s}(\mathbf{y}_k)}\!\bra{\mathbf{s}(\mathbf{y}_k)}.
\label{Eq:CycleInitialState}
\end{equation}
The input $u_k$ is encoded through the Hamiltonian local fields according to the rule $h_i(k)=m_i u_k$. The most general one-cycle evolution (with the state evolving from $\rho_{k,0}$ to $\rho_{k,f}$) is then
\begin{equation}
 \rho_{k,f}=\Phi_{u_k}\!\left[\rho_{k,0}\right],
\label{Eq:GeneralCycleMap}
\end{equation}
where $\Phi_{u_k}$ is a completely positive trace-preserving map determined by the annealing schedule, the input-dependent Ising Hamiltonian, and the environment. The experimentally accessible state is not the full density matrix but the set of coarse-grained magnetizations
\begin{equation}
 x_{k+1,\ell}
 =\frac{1}{g}\sum_{i\in\mathcal N_\ell}
 \Tr\!\left(\sigma_i^z\rho_{k,f}\right).
\label{Eq:GeneralReadout}
\end{equation}
With a finite number $M$ of shots, the ideal expectation value given by the Trace in Eq.~\eqref{Eq:GeneralReadout} is replaced by the sample mean presented in the main text. Equations~\eqref{Eq:ThresholdAgain}--\eqref{Eq:GeneralReadout} define the recurrent map independently of whether the dynamics inside a cycle are coherent, incoherent, or intermediate between the two.

A useful distinction should be kept in mind. The hardware does not take one microscopic measured bit string from shot $j$ and feed that same bit string into the next time step. Rather, $M$ independent shots are used to estimate the coarse-grained variables $\mathbf{x}_{k+1}$, and those averages are thresholded to prepare the next initialization. Thus the exact recurrence is a deterministic nonlinear map on the measured expectation values in the limit $M\rightarrow\infty$, supplemented by sampling noise at finite $M$. A stochastic matrix on microscopic bit strings is nevertheless the natural object for describing a single evaluation of that map.

\subsection{Coherent reverse annealing and the measurement kernel}

We first assume, for simplicity, that the system is closed during one cycle. This assumption will be relaxed in what follows. For a total annealing time $\Delta t$, the time-dependent Hamiltonian is
\begin{equation}
 H(t;u_k)=-\frac{A(t)}{2}\sum_{i=1}^{N}\sigma_i^x
 +\frac{B(t)}{2}\left[u_k\sum_i m_i\sigma_i^z+\sum_{i<j}J_{ij}\sigma_i^z\sigma_j^z\right]
\label{Eq:TimeDependentHamiltonian}
\end{equation}
whose corresponding propagator reads
\begin{equation}
 U_{u_k}(\Delta t)=\mathcal T\exp\!\left[-i\int_{0}^{\Delta t}\d t\,H(t;u_k)\right].
\end{equation}
The cycle map is therefore unitary:
\begin{equation}
 \Phi_{u_k}[\rho]=U_{u_k}\rho U_{u_k}^{\dagger}.
\label{Eq:CoherentCycle}
\end{equation}
Starting from a classical configuration $\ket{\mathbf{s}}$ and measuring at the end in the $\sigma^z$ basis gives the following coherent transition kernel
\begin{equation}
 \mathbb M^{\mathrm{coh}}_{\mathbf{s}\mathbf{s}'}(u_k)
 =\left|\bra{\mathbf{s}'}U_{u_k}\ket{\mathbf{s}}\right|^2.
\label{Eq:CoherentKernel}
\end{equation}
For every fixed initial configuration, unitarity implies
\begin{equation}
 \sum_{\mathbf{s}'}\mathbb M^{\mathrm{coh}}_{\mathbf{s}\mathbf{s}'}(u_k)=1,
\end{equation}
so the measured input--output relation is a row-stochastic kernel even though the dynamics before measurement are fully coherent. In fact, even if coherences are present during the cycle, the terminal measurement converts amplitudes into classical probabilities.

\subsection{Frozen-pause approximation and the Hamming graph}

To make the structure of Eq.~\eqref{Eq:CoherentKernel} explicit, we follow the short-time calculation developed for reverse annealing. During the pause, or during a sufficiently short segment of the ramps, the schedule may be treated as approximately constant. We write
\begin{equation}
 H_p=H_x+H_z,
\qquad
 H_x=-a\sum_i\sigma_i^x,
\qquad
 H_z=b\left(u_k\sum_i m_i\sigma_i^z+\sum_{i<j}J_{ij}\sigma_i^z\sigma_j^z\right),
\label{Eq:MicroFrozen}
\end{equation}
where, indicating with $s_p$ the pausing point in the schedule, $a=A(s_p)/2$ and $b=B(s_p)/2$. Every computational-basis state is an eigenstate of $H_z$,
\begin{equation}
 H_z\ket{\mathbf{s}}=E(\mathbf{s};u_k)\ket{\mathbf{s}},
\end{equation}
with
\begin{equation}
 E(\mathbf{s};u_k)
 =b\left(u_k\sum_i m_i s_i+\sum_{i<j}J_{ij}s_i s_j\right).
\label{Eq:MicroEnergy}
\end{equation}
The transverse term flips individual spins. If $\mathbf{s}^{(i)}$ denotes the configuration obtained from $\mathbf{s}$ by reversing spin $i$, then
\begin{equation}
 \sigma_i^x\ket{\mathbf{s}}=\ket{\mathbf{s}^{(i)}}.
\end{equation}
Consequently, the graph of directly connected computational-basis configurations is the $N$-dimensional Hamming graph, or hypercube. Its vertices are the $2^N$ bit strings, and an edge joins two strings when they differ in exactly one position. For $N=1,2,3$, this graph is respectively an edge, a square, and a cube, as shown in the following derivation.

\subsubsection{First-order Suzuki--Trotter expansion}

For a short time step $\Delta t$, the first-order product formula is
\begin{equation}
 e^{-i(H_x+H_z)\Delta t}
 =e^{-iH_x\Delta t}e^{-iH_z\Delta t}+O(\Delta t^2).
\label{Eq:FirstOrderTrotter}
\end{equation}
Applying the diagonal factor first gives
\begin{equation}
 e^{-iH_z\Delta t}\ket{\mathbf{s}}
 =e^{-iE(\mathbf{s};u_k)\Delta t}\ket{\mathbf{s}}.
\end{equation}
Since all $\sigma_i^x$ commute with one another,
\begin{equation}
 e^{-iH_x\Delta t}
 =\prod_{i=1}^{N}e^{ia\Delta t\sigma_i^x}
 =\prod_{i=1}^{N}\left[\cos(a\Delta t)\mathds 1+i\sin(a\Delta t)\sigma_i^x\right].
\label{Eq:TransverseFactorization}
\end{equation}
Expanding the trigonometric functions yields
\begin{equation}
 \cos(a\Delta t)=1-\frac{a^2\Delta t^2}{2}+O(\Delta t^4),
 \qquad
 \sin(a\Delta t)=a\Delta t+O(\Delta t^3).
\end{equation}
Keeping amplitudes through first order gives
\begin{equation}
 e^{-iH_p\Delta t}\ket{\mathbf{s}}
 =e^{-iE(\mathbf{s};u_k)\Delta t}
 \left[\ket{\mathbf{s}}+ia\Delta t\sum_i\ket{\mathbf{s}^{(i)}}\right]
 +O(\Delta t^2).
\label{Eq:MicroFirstOrder}
\end{equation}
The phase multiplying the entire state does not affect a measurement performed immediately after this single Trotter slice. The leading transition amplitude to each Hamming neighbor is $ia\Delta t$, and the associated probability is
\begin{equation}
 \Pr\bigl(\mathbf{s}\rightarrow\mathbf{s}^{(i)}\bigr)
 =a^2\Delta t^2+O(\Delta t^4).
\end{equation}
The probability of remaining in the original configuration follows either from Eq.~\eqref{Eq:TransverseFactorization} or from normalization,
\begin{equation}
 \Pr\bigl(\mathbf{s}\rightarrow\mathbf{s}\bigr)
 =1-Na^2\Delta t^2+O(\Delta t^4).
\end{equation}
Therefore, at this order, the observed dynamics are an unbiased local walk on the Hamming graph. The Ising energy appears only as a common phase and cannot yet bias the probabilities. This absence is not a physical statement about a finite pause; it is a consequence of truncating the short-time amplitudes before relative phases have had time to interfere.

\subsubsection{Second-order expansion and the first energy-dependent amplitudes}

To see where the energy landscape enters, one must retain the next order. We use the symmetric Strang splitting
\begin{equation}
 e^{-i(H_x+H_z)\Delta t}
 =e^{-iH_x\Delta t/2}e^{-iH_z\Delta t}e^{-iH_x\Delta t/2}
 +O(\Delta t^3).
\label{Eq:StrangSplitting}
\end{equation}
Expanding each exponential and multiplying the factors gives
\begin{equation}
 e^{-iH_p\Delta t}
 =\mathds 1-i(H_x+H_z)\Delta t
 -\frac{\Delta t^2}{2}\left(H_x^2+H_z^2+H_xH_z+H_zH_x\right)
 +O(\Delta t^3),
\label{Eq:SecondOrderOperator}
\end{equation}
which agrees with the direct Taylor expansion of $e^{-iH_p\Delta t}$. Acting on $\ket{\mathbf{s}}$, the required terms are
\begin{align}
 H_z\ket{\mathbf{s}}&=E_{\mathbf{s}}\ket{\mathbf{s}},\\
 H_x\ket{\mathbf{s}}&=-a\sum_i\ket{\mathbf{s}^{(i)}},\\
 H_x^2\ket{\mathbf{s}}&=a^2\left[N\ket{\mathbf{s}}+2\sum_{i<j}\ket{\mathbf{s}^{(ij)}}\right],
\label{Eq:HxSquared}
\end{align}
where $E_{\mathbf{s}}\equiv E(\mathbf{s};u_k)$ and $\mathbf{s}^{(ij)}$ differs from $\mathbf{s}$ at spins $i$ and $j$. Furthermore,
\begin{align}
 H_xH_z\ket{\mathbf{s}}&=-aE_{\mathbf{s}}\sum_i\ket{\mathbf{s}^{(i)}},\\
 H_zH_x\ket{\mathbf{s}}&=-a\sum_iE_{\mathbf{s}^{(i)}}\ket{\mathbf{s}^{(i)}}.
\end{align}
Substitution into Eq.~\eqref{Eq:SecondOrderOperator} yields
\begin{align}
 e^{-iH_p\Delta t}\ket{\mathbf{s}}
 ={}&\left[1-iE_{\mathbf{s}}\Delta t
 -\frac{\Delta t^2}{2}\left(E_{\mathbf{s}}^2+Na^2\right)\right]\ket{\mathbf{s}}
 \nonumber\\
 &+\sum_i\left[ia\Delta t
 +\frac{a\Delta t^2}{2}\left(E_{\mathbf{s}}+E_{\mathbf{s}^{(i)}}\right)\right]\ket{\mathbf{s}^{(i)}}
 \nonumber\\
 &-a^2\Delta t^2\sum_{i<j}\ket{\mathbf{s}^{(ij)}}+O(\Delta t^3).
\label{Eq:SecondOrderState}
\end{align}
This equation makes the hierarchy of paths transparent. Zero flips contribute to the survival amplitude, one transverse action reaches a nearest neighbor, and two transverse actions either return to the original vertex or reach a configuration at Hamming distance two. The Ising energy first enters the one-flip amplitude through the sum $E_{\mathbf{s}}+E_{\mathbf{s}^{(i)}}$.

At the level of probabilities, however, the real second-order correction in the one-flip amplitude does not interfere with the purely imaginary first-order amplitude. Thus
\begin{equation}
 \left|\bra{\mathbf{s}^{(i)}}e^{-iH_p\Delta t}\ket{\mathbf{s}}\right|^2= \left|ia\Delta t+\frac{a\Delta t^2}{2}(E_{\mathbf{s}}+E_{\mathbf{s}^{(i)}})\right|^2
 =a^2\Delta t^2+O(\Delta t^4).
\end{equation}
Energy-dependent corrections to a single-slice transition probability therefore arise at higher order than $\Delta t^2$. This observation resolves an apparent tension in the short-time calculation: the energy landscape is already present in the amplitudes at second order, but a measurable bias requires either higher-order interference, several successive slices, or evolution over a finite hold time. The latter is the physically relevant situation for the device.

It is useful to express the relevant energy difference explicitly. Flipping spin $i$ changes the diagonal energy by
\begin{equation}
 \Delta E_i(\mathbf{s};u_k)
 \equiv E(\mathbf{s}^{(i)};u_k)-E(\mathbf{s};u_k)
 =-2b s_i\left(m_i u_k+\sum_{j\neq i}J_{ij}s_j\right).
\label{Eq:EnergyDifferenceCoherent}
\end{equation}
Thus the input and the neighboring spins appear only through the local effective longitudinal field
\begin{equation}
 \epsilon_i(\mathbf{s};u_k)
 =b\left(m_i u_k+\sum_{j\neq i}J_{ij}s_j\right),
 \qquad
 \Delta E_i=-2s_i\epsilon_i.
\label{Eq:LocalDetuning}
\end{equation}

\subsubsection{Finite-time local dynamics}

A simple resummation of the short-time expansion is obtained by freezing the neighboring spins during a local transition. The effective two-level Hamiltonian for spin $i$ is then
\begin{equation}
 H_i(\mathbf{s})=-a\sigma_i^x+\epsilon_i(\mathbf{s};u_k)\sigma_i^z.
\label{Eq:MicroLocalH}
\end{equation}
Since $H_i^2=(a^2+\epsilon_i^2)\mathds 1$, its exponential is exact,
\begin{equation}
 e^{-iH_i\Delta t}
 =\cos(\Omega_i\Delta t)\mathds 1
 -i\frac{\sin(\Omega_i\Delta t)}{\Omega_i}
 \left(-a\sigma_i^x+\epsilon_i\sigma_i^z\right),
 \qquad
 \Omega_i=\sqrt{a^2+\epsilon_i^2}.
\end{equation}
Starting from a $\sigma_i^z$ eigenstate, the probability of observing the opposite spin is
\begin{equation}
 p_{i,\mathrm{coh}}^{\mathrm{flip}}(\mathbf{s})
 =\frac{a^2}{a^2+\epsilon_i(\mathbf{s};u_k)^2}
 \sin^2\!\left(\sqrt{a^2+\epsilon_i(\mathbf{s};u_k)^2}\,\Delta t\right).
\label{Eq:MicroRabi}
\end{equation}
Equation~\eqref{Eq:MicroRabi} displays directly the bias that is invisible in the leading short-time probability. The longitudinal field suppresses the oscillation amplitude and changes its frequency. Because $\epsilon_i$ depends on both the input and the neighboring spins, the coherent transition kernel is nonlinear and nonfactorizable whenever $J_{ij}\neq0$. The frozen-neighbor approximation is exact at $J_{ij}=0$ and serves as a transparent local approximation when the neighbors evolve more slowly than the selected spin.

\subsection{General open-system reverse annealing}

We now relax the assumption that the quantum annealer is coherent. In fact, real quantum annealers are coupled to an environment, and the reverse-annealing time may be comparable with relaxation and dephasing times. The natural replacement for Eq.~\eqref{Eq:CoherentCycle} is therefore a quantum dynamical map generated by a time-dependent Liouvillian,
\begin{equation}
 \frac{\d}{\d t}\rho=\mathcal L_{u_k}(t)[\rho],
 \qquad
 \Phi_{u_k}=\mathcal T\exp\!\left[\int_0^{\Delta t}\d t\,\mathcal L_{u_k}(t)\right].
\label{Eq:LiouvillianCycle}
\end{equation}
In a weak-coupling Markovian description, one writes
\begin{align}
 \mathcal L_{u_k}(t)[\rho]
 =-i\left[H(t;u_k)+H_{\mathrm{LS}}(t),\rho\right] +\sum_{\alpha,\omega}\gamma_{\alpha}(\omega)
 \left[L_{\alpha,\omega}(t)\rho L_{\alpha,\omega}^{\dagger}(t)
 -\frac{1}{2}\left\{L_{\alpha,\omega}^{\dagger}(t)L_{\alpha,\omega}(t),\rho\right\}\right].
\label{Eq:OpenSystemGenerator}
\end{align}
Here $H_{\mathrm{LS}}$ is the Lamb-shift Hamiltonian, $\{\gamma_{\alpha}(\omega)\}_{\alpha}$ are the damping rates related to the Bohr frequency $\omega$ and $\{L_{\alpha,\omega}\}_{\alpha}$ are the corresponding jump operators. If $S_\alpha$ denotes the system operator coupled to bath channel $\alpha$, then 
\begin{equation}
 L_{\alpha,\omega}(t)
 =\sum_{\epsilon_b(t)-\epsilon_a(t)=\omega}
 \Pi_a(t)S_\alpha\Pi_b(t),
 \qquad
 \Pi_a(t)=\ket{\epsilon_a(t)}\!\bra{\epsilon_a(t)},
\label{Eq:JumpOperators}
\end{equation}
where $\ket{\epsilon_a(t)}$ is one instantaneous eigenvector of the time-dependent Hamiltonian. 

For a thermal environment, upward and downward rates satisfy the Kubo--Martin--Schwinger relation
\begin{equation}
 \gamma_\alpha(-\omega)=e^{-\beta\omega}\gamma_\alpha(\omega),
\label{Eq:KMSRates}
\end{equation}
with $e^{-\beta\omega}$ denoting the Boltzmann factor determined by the environment temperature

This formulation is closely related to the open-system adiabatic framework developed by Zanardi, Lidar, and collaborators~\cite{ZanardiLidarOpenAdiabatic,AlbashLidarReview}. In that setting, adiabatic transport is formulated for spectral subspaces of the Liouvillian. The slowly followed object may be a mixed instantaneous steady state or, more generally, the zero-eigenvalue sector of the Liouvillian, rather than a pure eigenstate of the Hamiltonian.

The basis in which decoherence is most naturally described depends on the physical regime. In a weak-coupling treatment in which bath correlations are short compared with the annealing timescale, the instantaneous energy basis is often appropriate. In a singular-coupling or very rapid dephasing regime, the computational basis may be more natural. We do not need to select one of these limits in order to define the reservoir. For any completely positive map, the corresponding kernel is
\begin{equation}
 \mathbb M_{\mathbf{s}\mathbf{s}'}(u_k)
 =\Tr\!\left[\Pi_{\mathbf{s}'}\Phi_{u_k}[\Pi_{\mathbf{s}}]\right],
 \qquad
 \Pi_{\mathbf{s}}=\ket{\mathbf{s}}\!\bra{\mathbf{s}}.
\label{Eq:OpenKernel}
\end{equation}
Positivity of $\Phi_{u_k}$ implies $\mathbb M_{\mathbf{s}\mathbf{s}'}\ge0$, while trace preservation gives
\begin{equation}
 \sum_{\mathbf{s}'}\mathbb M_{\mathbf{s}\mathbf{s}'}(u_k)=1.
\end{equation}
Equation~\eqref{Eq:OpenKernel} is therefore the open-system generalization of Eq.~\eqref{Eq:CoherentKernel}. It includes coherent, partially coherent, and fully incoherent dynamics in a single expression.

\subsection{Rapid-dephasing limit and the Pauli master equation}

In the fully classical regime, the time evolution of the state can be replaced by the time evolution of classical state probability distribution. To see this, suppose now that dephasing removes off-diagonal density-matrix elements on a timescale that is short compared with the population dynamics. For the equation
\begin{equation}
 P_{\mathbf{s}}(t)=\bra{\mathbf{s}}\rho(t)\ket{\mathbf{s}},
\end{equation}
the coherences may be adiabatically eliminated, leaving a classical Pauli master equation,
\begin{equation}
 \frac{\d}{\d t} P_{\mathbf{s}}
 =\sum_{\mathbf{s}'\neq\mathbf{s}}
 \left[W_{\mathbf{s}'\rightarrow\mathbf{s}}(t;u_k)P_{\mathbf{s}'}
 -W_{\mathbf{s}\rightarrow\mathbf{s}'}(t;u_k)P_{\mathbf{s}}\right],
\label{Eq:PauliMaster}
\end{equation}
where $W_{\mathbf{s}'\rightarrow\mathbf{s}}$ denotes the transition rate from state $\mathbf{s}'$ to state $\mathbf{s}$.

Using row-vector convention, this can be written as
\begin{equation}
 \dot{\mathbf P}(t)=\mathbb W(t;u_k)\mathbf P(t),
\label{Eq:PopulationGenerator}
\end{equation}
where $\mathbf P$ is the row vector containing the probabilities of all classical configurations and $\mathbb W$ is the corresponding transition-rate matrix.

The off-diagonal entries of $\mathbb W$ are the transition rates, and probability conservation fixes the diagonal entries,
\begin{equation}
 \mathbb W_{\mathbf{s}\mathbf{s}'}=W_{\mathbf{s}\rightarrow\mathbf{s}'},
 \quad \mathbf{s}'\neq\mathbf{s},
 \qquad
 \mathbb W_{\mathbf{s}\mathbf{s}}=-\sum_{\mathbf{r}\neq\mathbf{s}}W_{\mathbf{s}\rightarrow\mathbf{r}}.
\end{equation}
The corresponding propagator is
\begin{equation}
 \mathbb M_{\mathrm{inc}}(u_k)
 =\mathcal T\exp\!\left[\int_0^{\Delta t}\d t\,\mathbb W(t;u_k)\right].
\label{Eq:IncoherentKernel}
\end{equation}
Thus the coherent amplitude kernel is replaced by an ordinary kinetic propagator.

For local processes, the dominant transitions connect Hamming neighbors $\mathbf{s}$ and $\mathbf{s}^{(i)}$. Their energy difference is
\begin{align}
\Delta E_i(\mathbf{s};u_k) \equiv E(\mathbf{s}^{(i)};u_k)-E(\mathbf{s};u_k) =-s_iB(t)\left(m_i u_k+\sum_{j\neq i}J_{ij}s_j\right).
\label{Eq:EnergyDifference}
\end{align}
The rate can therefore be written in the generic form
\begin{equation}
 \Gamma_i\!\left(t,\Delta E_i(\mathbf{s};u_k)\right) \equiv W_{\mathbf{s}\rightarrow\mathbf{s}^{(i)}}(t;u_k).
\end{equation}
For a thermal bath, microscopic reversibility implies local detailed balance,
\begin{equation}
 \frac{W_{\mathbf{s}\rightarrow\mathbf{s}^{(i)}}}
 {W_{\mathbf{s}^{(i)}\rightarrow\mathbf{s}}}
 =e^{-\beta\Delta E_i(\mathbf{s};u_k)}.
\label{Eq:DetailedBalance}
\end{equation}
One phenomenological choice satisfying Eq.~\eqref{Eq:DetailedBalance} is the heat-bath or Glauber rate
\begin{equation}
 W_{\mathbf{s}\rightarrow\mathbf{s}^{(i)}}
 =\frac{\kappa_i(t)}{1+e^{\beta\Delta E_i(\mathbf{s};u_k)}},
\label{Eq:GlauberRate}
\end{equation}
where $\kappa_i(t)$ is a generic positive function.

Indeed, replacing $\Delta E_i$ by $-\Delta E_i$ for the reverse move immediately gives the ratio in Eq.~\eqref{Eq:DetailedBalance}. The amenability of the incoherent regime to Monte Carlo sampling was recently harnessed in \cite{Sathe2026ClassicalCriticality}.

A rate generated by incoherent tunneling in the presence of strong dephasing has a different prefactor but the same configuration dependence. Schematically,
\begin{equation}
 W_{\mathbf{s}\rightarrow\mathbf{s}^{(i)}}
 \simeq
 \frac{A(t)^2\Gamma_{\varphi,i}(t)}
 {\Gamma_{\varphi,i}(t)^2+\Delta E_i(\mathbf{s};u_k)^2}
 f_\beta\!\left(\Delta E_i\right),
\label{Eq:IncoherentTunnelingRate}
\end{equation}
where $\Gamma_{\varphi,i}$ is a dephasing scale and the thermal factor $f_\beta$ is chosen so that Eq.~\eqref{Eq:DetailedBalance} holds. The Lorentzian factor has a transparent origin: dephasing broadens the energy-conservation condition, while the transverse field supplies the matrix element for a spin flip. Equation~\eqref{Eq:IncoherentTunnelingRate} is the kinetic analogue of the coherent Rabi probability in Eq.~\eqref{Eq:MicroRabi}. Oscillatory population transfer is replaced by an irreversible transition rate.

\subsection{Memory and non-Markovianity}

We emphasize that, in all the regimes described here, the reservoir state consists of a set of real-valued variables, $\mathbf{x}_{k}$, while the annealer dynamics provide the evolution rule used to update their values. The system’s memory is therefore stored in these classical variables rather than in the quantum state.

As already mentioned, for the reservoir to operate properly, it must satisfy the echo state property. This requires the reservoir dynamics to progressively erase information about the reservoir's initial condition and, consequently, to be contractive. In the reservoir architecture described here, contractivity arises from the procedure used to update the reservoir state at each time step. In particular, the coarse-grained measurement of the magnetizations, together with the preparation of the initial states according to the proposed threshold rule, ensures that this property is satisfied. Therefore, in our architecture, which relies on classical memory, the echo state property can also be achieved when the quantum annealer operates in a closed-system regime. By contrast, as discussed in Ref.~\cite{Sannia2024}, closed quantum systems are unsuitable as reservoirs when information is stored over time directly in the quantum states.

Moreover, in the absence of non-Markovian effects, the memory of any reservoir tends to decay exponentially with time, regardless of whether the information is stored in a quantum or a classical state~\cite{Sannia2026}. As shown in the main text, this limitation can be overcome through the use of delayed feedback. Such feedback enables memory revivals, producing a richer and less trivial temporal memory structure than a simple monotonic decay, as reflected in the performance results reported in the main text.

\section{Why the non-interacting case fails}
\label{sec:noninteracting-proof}

As experimentally shown in the main text, setting $J_{ij}=0$ removes the useful temporal-processing capability of the reservoir. In the following, we provide an analytical proof of this behaviour.

The main idea of the proof is that a factorized noninteracting reservoir has only two possibilities. The first is that the local thresholded branch retains information about its initial condition, in which case the echo-state property fails. An alternative possibility is that the local channels contract and erase that distinction, in which case the reservoir satisfies the echo-state property, but no memory of past inputs can be stored.

The echo-state property requires that, for any fixed input sequence, two reservoir trajectories that are initialized differently will asymptotically converge~\cite{Yildiz2012}.  Denoting by $\mathcal F_k$ the complete update at step $k$, for any pair of initial reservoir states $\mathbf{x}_0$ and $\bar{\mathbf{x}}_0$, the echo-state property formally translates to the condition

\begin{equation}
\label{Eq:EchoState}
\lim_{T\rightarrow\infty}
\left\|
\mathcal F_{T-1}\circ\cdots\circ\mathcal F_0(\mathbf x_0)
-
\mathcal F_{T-1}\circ\cdots\circ\mathcal F_0(\bar{\mathbf x}_0)
\right\|=0,
\end{equation}
where the input sequence is identical in the two evolutions, and only the initial reservoir states differ.

\subsection{Factorization of the noninteracting maps}

When the couplings vanish, the Hamiltonian is a sum of local terms,
\begin{equation}
H_k(t)=\sum_{i=1}^{N}H_k^{(i)}(t),
\qquad
H_k^{(i)}(t)=-\frac{A(t)}{2}\sigma_i^x+\frac{B(t)}{2}m_i u_k\sigma_i^z.
\label{Eq:LocalHamiltonianNonInt}
\end{equation}
For a closed system, terms acting on different qubits commute, so the corresponding unitary evolution factorizes into single-qubit operators $U_{u_k}^{(i)}$:
\begin{equation}
U_{u_k}=\bigotimes_{i=1}^{N}U_{u_k}^{(i)}.
\label{Eq:FactorizedUnitary}
\end{equation}
More generally, if the system--bath coupling is also local and does not mediate interactions between different qubits, the annealer evolution can be written according to the product form
\begin{equation}
\Phi_{u_k}=\bigotimes_{i=1}^{N}\Phi_{i,u_k},
\label{Eq:FactorizedChannel}
\end{equation}
where $\Phi_{i,u_k}$ is an input-dependent quantum channel acting on the $i$-th qubit. 

In the rapid-dephasing limit, Eq.~\eqref{Eq:FactorizedChannel} reduces to
\begin{equation}
\mathbb M_{\mathrm{inc}}(u_k)
=\bigotimes_{i=1}^{N}\mathbb M_{i}(u_k),
\label{Eq:FactorizedClassicalKernel}
\end{equation}
where $\mathbb M_{i}(u_k)$ is a two-state stochastic kernel.

Equations~\eqref{Eq:FactorizedUnitary}--\eqref{Eq:FactorizedClassicalKernel} are the common mathematical content of the noninteracting limit. Because the thresholding rule prepares a computational-basis state, the local input to qubit $i$ at cycle $k$ is
\begin{equation}
\rho_{i,k}^{\mathrm{in}}
=\ket{s_{i,k}}\!\bra{s_{i,k}},
\qquad
s_{i,k}=2z_{i,k}-1,
\qquad
z_{i,k}=\Theta\!\left(y_{k,\ell(i)}-\theta_i\right).
\label{Eq:LocalPreparedState}
\end{equation}
For any qubit, the two possible measured outputs are
\begin{equation}
\mu_{i,k}^{(+1)}
=\Tr\!\left[\sigma_i^z\Phi_{i,u_k}(\ket{+1}\!\bra{+1})\right],
\qquad
\mu_{i,k}^{(-1)}
=\Tr\!\left[\sigma_i^z\Phi_{i,u_k}(\ket{-1}\!\bra{-1})\right].
\label{Eq:TwoLocalOutputs}
\end{equation}
Since $s_{i,k}=\pm1$, the most general local update can always be written in affine form,
\begin{equation}
 x_{i,k+1}=c_{i,k}+a_{i,k}s_{i,k}
 =c_{i,k}+a_{i,k}\left(2z_{i,k}-1\right),
\label{Eq:GenericNonInteractingMap}
\end{equation}
with
\begin{equation}
 c_{i,k}=\frac{\mu_{i,k}^{(+1)}+\mu_{i,k}^{(-1)}}{2},
 \qquad
 a_{i,k}=\frac{\mu_{i,k}^{(+1)}-\mu_{i,k}^{(-1)}}{2}.
\label{Eq:AffineCoefficients}
\end{equation}
The coefficient $a_{i,k}$ measures how strongly the output distinguishes the two prepared branches, while $c_{i,k}$ is a bias which depends on the specific properties of $\Phi_{i,u_k}$.

Equation~\eqref{Eq:GenericNonInteractingMap} also contains the classical population dynamics.  If $\mathbb M_i(u_k)$ is a row-stochastic two-state kernel, then
\begin{equation}
 \mu_{i,k}^{(s)}
 =\sum_{s'=\pm1}s'\,
 \mathbb M_i(u_k)_{s,s'},
\label{Eq:ClassicalLocalMean}
\end{equation}
which, inserted into Eq.~\eqref{Eq:AffineCoefficients}, gives exactly the same affine recurrence. 

\subsection{Threshold branches and the echo-state property}

The thresholded scalar map has two possible outputs at every step,
\begin{equation}
 q_{i,k}^{(+)}=c_{i,k}+a_{i,k},
 \qquad
 q_{i,k}^{(-)}=c_{i,k}-a_{i,k}.
\label{Eq:TwoBranchesGeneric}
\end{equation}
The convergence of two trajectories is determined by the positions of these two values relative to the threshold $\theta_i$.  Suppose first that they lie on opposite sides:
\begin{equation}
 \bigl(q_{i,k}^{(+)}-\theta_i\bigr)
 \bigl(q_{i,k}^{(-)}-\theta_i\bigr)<0.
\label{Eq:StraddlingCondition}
\end{equation}
Then the threshold distinguishes the two branch labels.  If two trajectories enter step $k$ with opposite labels, they leave it with distinct outputs and are thresholded into opposite labels again.  If Eq.~\eqref{Eq:StraddlingCondition} holds for any $k$, the branch distinction is invariant. Their separation after the cycle is
\begin{equation}
 \left|q_{i,k}^{(+)}-q_{i,k}^{(-)}\right|=2|a_{i,k}|.
\end{equation}
Consequently, if there is an $a_*>0$ such that $|a_{i,k}|\ge a_*$ along an infinite tail of the input sequence, then
\begin{equation}
 \lim_{k\rightarrow\infty}\sup
 \left|x_{i,k}-\bar x_{i,k}\right|\ge 2a_*>0,
\end{equation}
and the echo-state property in Eq.~\eqref{Eq:EchoState} fails.  

Indeed, inputs satisfying Eq.~\eqref{Eq:StraddlingCondition} cannot be reliably processed by a reservoir-computing algorithm. Even when such inputs are embedded in a sequence that converges because of other inputs exhibiting the opposite behavior, they can still hinder the training process. In particular, inputs satisfying Eq.~\eqref{Eq:StraddlingCondition} dynamically modify the reservoir’s convergence rate, thereby making its response explicitly time-dependent. Since the linear optimization used to train the reservoir assumes a time-independent input–output mapping, this class of inputs must be avoided.

For an acceptable input sequence, all the outputs lie on the same side of the threshold:
\begin{equation}
 \bigl(q_{i,k}^{(+)}-\theta_i\bigr)
 \bigl(q_{i,k}^{(-)}-\theta_i\bigr)>0.
\label{Eq:MergingCondition}
\end{equation}

In this case, any two initially distinct reservoir states converge to the same state after a single time step. Therefore, the convergence time required for the echo-state property to hold is exactly one iteration. This implies that the reservoir can retain information for, at most, one time step. Consequently, information about the preceding input sequence is immediately erased. In this regime, the satisfaction of the echo-state property therefore coincides with a complete loss of memory. Moreover, for a randomly configured reservoir driven by a random input sequence, this condition is highly unlikely to be satisfied. Indeed, our experimental results show that, for the models studied in the main text, the noninteracting regime exhibits no memory.

We emphasize that increasing the number of independent qubits does not resolve this issue. Indeed, any linear combination of memoryless outputs remains memoryless. Likewise, introducing delayed feedback does not change this conclusion, because the thresholding rule immediately erases the information carried by the feedback signal.

Finally, we note that the preceding arguments no longer apply in the interacting regime, where $J_{i,j}\neq 0$, because many-body interactions give rise to substantially more complex dynamics. In this case, increasing the number of qubits can generate collective effects that modify the impact of the thresholding procedure used to prepare the spin states. In particular, if only a subset of qubits properly converged, the interactions can propagate the retained information throughout the rest of the network over successive iterations. As a result, the state of the entire reservoir is expected to converge to a nontrivial, well-defined function of the recent input history.

\begin{figure}[t!]
\centering
\IfFileExists{figures/Ideal_perfs.pdf}{%
\includegraphics[width=0.7\linewidth,keepaspectratio]{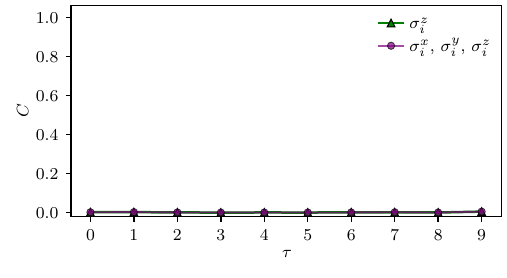}%
}{%
\fbox{\rule{0pt}{4cm}\rule{0.48\linewidth}{0pt}}%
}
\caption{Linear memory capacity as a function of the delay $\tau$, obtained from ideal numerical simulations of the small-scale coherent model. The green curve uses an output layer built from $\langle\sigma_i^z\rangle$. The purple curve additionally includes $\langle\sigma_i^x\rangle$ and $\langle\sigma_i^y\rangle$. The feedback delay is $d=1$.}
\label{fig:IdealPerf}
\end{figure}

\section{Small-scale ideal coherent model}
\label{sec:small-scale-ideal}

The preceding proof concerns the exact structural properties of the noninteracting coherent limit. We now return to the interacting model and ask a different question: whether a small, ideal, closed-system reservoir already exhibits the temporal processing observed on the large device. This numerical experiment is not intended as a faithful noise model of the hardware. Rather, it isolates the coherent Hamiltonian mechanism derived in Sec.~\ref{sec:micro} and removes finite-shot noise, calibration drift, relaxation, and dephasing.

In the main text, the measured memory capacity improves monotonically as the number of physical qubits increases. To complement that observation, we simulate a reservoir of five interacting qubits. Its Hilbert-space dimension is only $2^5=32$, so the full time-dependent Schr\"odinger equation can be integrated without approximation,
\begin{equation}
i\frac{\d}{\d t}\ket{\psi_k(t)}=H(t;u_k)\ket{\psi_k(t)},
\qquad
\ket{\psi_k(0)}=\ket{\mathbf s(\mathbf y_k)}.
\label{Eq:IdealSchrodinger}
\end{equation}
The Hamiltonian $H(t;u_k)$ is the reverse-annealing Hamiltonian of Eq.~\eqref{Eq:TimeDependentHamiltonian}, with the same input encoding and thresholded initialization as in the experimental protocol.

The discretized scheduling functions $A(t)$ and $B(t)$ are obtained from the D-Wave repository~\cite{DWaveOceanSDK}. Because the solver evaluates the Hamiltonian at intermediate times, the tabulated schedule is converted into continuous functions using shape-preserving piecewise-cubic interpolation. This choice avoids the artificial oscillations that can be introduced by an ordinary high-order spline while preserving the monotonic structure of the hardware calibration curves. The resulting time-dependent problem is integrated with the QuTiP solver~\cite{qutip5}.

At the end of cycle $k$, the ideal local observables are evaluated directly from the pure state,
\begin{equation}
x_{i,k+1}^{(z)}
=\bra{\psi_k(t_f)}\sigma_i^z\ket{\psi_k(t_f)}.
\label{Eq:IdealZReadout}
\end{equation}
The coarse-grained reservoir variables are obtained by averaging these quantities over the chosen groups of qubits, exactly as in Eq.~\eqref{Eq:GeneralReadout}. Unlike the hardware experiment, no projective samples are drawn: the expectation values are known exactly. The simulation therefore represents the limit $M\rightarrow\infty$ and contains no statistical shot noise.

This distinction is useful but should not be interpreted as making the simulation uniformly more favorable. Shot noise is removed, but so are dissipative effects that may assist contraction and exploration of the energy landscape. The ideal calculation tests whether coherent dynamics alone, at this small system size and with this readout, supply a useful balance between fading memory and state separation. It does not test the full open-system mechanism available to the physical annealer.

The training and testing protocol is kept identical to that used in the main text. A random input stream drives the system, an initial washout interval is discarded, and a linear readout is trained to reconstruct delayed inputs. For each delay $\tau$, the target is
\begin{equation}
\hat o_k=u_{k-\tau},
\end{equation}
and the memory capacity is the squared correlation between the target and the trained prediction. Figure~\ref{fig:IdealPerf} shows the result for one representative realization. The capacity remains close to zero for every tested delay, indicating that the five-qubit coherent reservoir does not retain the input history in a form accessible to the linear output layer.

One possible explanation would be that the information is present in quantum coherences but is invisible to a $\sigma^z$ readout. We test this possibility by enlarging the feature vector to include all three local Bloch components,
\begin{equation}
\mathbf x_k=
\left(
\langle\sigma_1^x\rangle,\ldots,\langle\sigma_N^x\rangle,
\langle\sigma_1^y\rangle,\ldots,\langle\sigma_N^y\rangle,
\langle\sigma_1^z\rangle,\ldots,\langle\sigma_N^z\rangle
\right)_k.
\label{Eq:FullBlochReadout}
\end{equation}
This triples the number of local observables and includes quantities unavailable in the standard D-Wave readout. Nevertheless, the capacity curve is essentially unchanged. The failure is therefore not simply due to omitting the transverse components of the final state.

The ideal simulation and the analytical proof make complementary statements. The proof shows that the coherent noninteracting model is structurally unsuitable: its factorized threshold dynamics either violate the echo-state property or collapse trivially. The simulation retains interactions but shows that five coherent qubits still provide too small a feature space, or too weak a balance of mixing and memory, to solve the temporal benchmark. Together with the experimental scaling results, this supports the conclusion that large system size and interactions are both important. At the same time, the open-system derivation cautions against interpreting the small coherent simulation as a model of the large hardware device: incoherent tunneling, thermal relaxation, and dissipation-assisted contraction may contribute substantially to the reservoir map realized experimentally.

\section{Classical benchmark: spin-vector Monte Carlo}

\begin{figure}[t!]
\centering
{%
\includegraphics[width=0.7\linewidth,keepaspectratio]{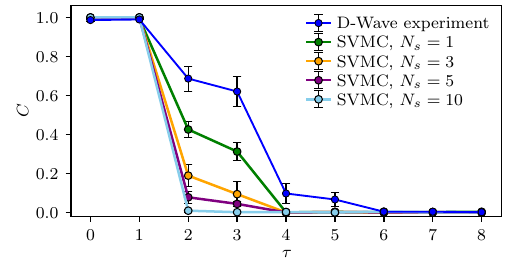}%
}
\caption{Comparison of the capacity values at varying delay ($\tau$) obtained from SVMC simulations versus the experimental D-Wave results presented in the main text. The considered delayed feedback is $d=1$, and the number of outputs is $K=45$. SVMC outputs were calculated using $10^3$ trajectories, equivalent to the $M=10^3$ measurements employed in the experiments presented in the main text. Consistent with the main text, error bars indicate a standard deviation over five independent realizations of the couplings, masks, and input data. SVMC was implemented using the same realizations as those employed in the experiments presented in the main text, while the training and testing procedures follow the same strategy adopted there for the real experiments.}
\label{fig:SVMC}
\end{figure}

To investigate the role of quantum effects in the reservoir architecture considered in the main text, we employ the spin-vector Monte Carlo (SVMC) model as a natural classical benchmark. This scheme is commonly used as a semiclassical surrogate for transverse-field Ising annealing~\cite{ShinSVMC,BoixoExperimentalSignature,AlbashLidarReview}. It represents a stringent benchmark because it retains the interacting energy landscape, thermal stochasticity, and a schedule-dependent transverse contribution, while discarding entanglement, coherent phases, and genuine quantum tunneling between macroscopically distinct configurations.  Each qubit is replaced by a classical unit vector restricted to the $x$--$z$ plane,
\begin{equation}
 \mathbf n_i=(\sin\theta_i,0,\cos\theta_i).
\end{equation}
The replacements
\begin{equation}
 \sigma_i^x\longrightarrow\sin\theta_i,
 \qquad
 \sigma_i^z\longrightarrow\cos\theta_i
\end{equation}
turn the considered transverse-field Ising Hamiltonian into the classical energy function
\begin{align}
 \mathcal H_{\mathrm{SVMC}}(t;u_k,\boldsymbol\theta)
 =-\frac{A(t)}{2}\sum_i\sin\theta_i
 +\frac{B(t)}{2}\left[u_k\sum_i m_i\cos\theta_i
 +\sum_{i<j}J_{ij}\cos\theta_i\cos\theta_j\right].
\label{Eq:SVMCEnergy}
\end{align}
The reverse anneal is simulated by discretizing the annealing schedule into ten equally spaced points along the annealing trajectory, at which the hardware controls $A(t)$ and $B(t)$ are evaluated. At each point, a number of $N_s$ Monte Carlo sweeps is performed, with the spins visited once per sweep in a randomized order. For each spin $i$, the considered Monte Carlo scheme proposes a new angle uniformly sampled from the set $[0,\pi]$:
\begin{equation}
    \theta_i \longrightarrow \theta_i',
    \qquad
    \theta_i' \sim \mathcal{U}(0,\pi).
\end{equation}
 The proposed move is then accepted or rejected according to the Metropolis criterion. Given the corresponding energy change 
\begin{equation}
 \Delta\mathcal H_i
 =\mathcal H_{\mathrm{SVMC}}(\theta_i')-
 \mathcal H_{\mathrm{SVMC}}(\theta_i),
\end{equation}
 the Metropolis update is accepted with probability
\begin{equation}
 p_{\mathrm{acc}}
 =\min\left\{1,e^{-\beta\Delta\mathcal H_i}\right\},
\label{Eq:SVMCMetropolis}
\end{equation}
where, taking the nominal operating temperature of \texttt{Advantage2\_system1} to be
$T \simeq 20\,\mathrm{mK}$, and expressing the thermal energy in units
consistent with the publicly available D-Wave annealing functions $A(t)$
and $B(t)$, given in GHz, we set
\begin{equation}
    \frac{k_B T}{h} \simeq 0.4167\,\mathrm{GHz}.
\end{equation}

Repeating these local moves for a prescribed number of sweeps at each schedule point defines a classical stochastic dynamics where the number of sweeps, $N_s$, plays a role analogous to the amount of relaxation time available to the physical system.

 The initialization of each reverse-annealing cycle is obtained by representing the binary spin states $s_i=\pm1$ as angular variables, with $s_i=+1$ corresponding to $\theta_i=0$ and $s_i=-1$ corresponding to $\theta_i=\pi$. Following the reservoir architecture presented in the main text, the input is inserted through the same masks $m_i$, and the delayed-feedback thresholding is applied without modification. At the end of a cycle the SVMC readout is
\begin{equation}
 x_{k+1,\ell}^{\mathrm{SVMC}}
 =\frac{1}{g}\sum_{i\in\mathcal N_\ell}
 \left\langle\cos\theta_i\right\rangle,
\label{Eq:SVMCReadout}
\end{equation}
where the average is taken over independent Monte Carlo trajectories. The effect of finite-shot noise was reproduced by drawing a finite ensemble of trajectories with the same size $M$ used in the experiment.

In Figure~\ref{fig:SVMC}, the linear memory capacity computed with SVMC simulations, varying the number of sweeps $N_s$, is compared with the one obtained from the experiments performed on the D-Wave machine from the main text. The considered comparison matches the programmed graph, couplings, temperature, masks, and readout statistics. Moreover, the training procedure and the evaluation protocol used for this SVMC-based reservoir are the same as those described in the main text.

The systematic discrepancies observed show that the reservoir dynamical features are not fully captured by the spin-vector description. This result therefore suggests that quantum effects play a crucial role in the performance of the quantum reservoir model studied.

\section{Details on the chaotic time-series forecasting}

\begin{figure}[t!]
    \centering  \includegraphics[width=\linewidth, keepaspectratio]{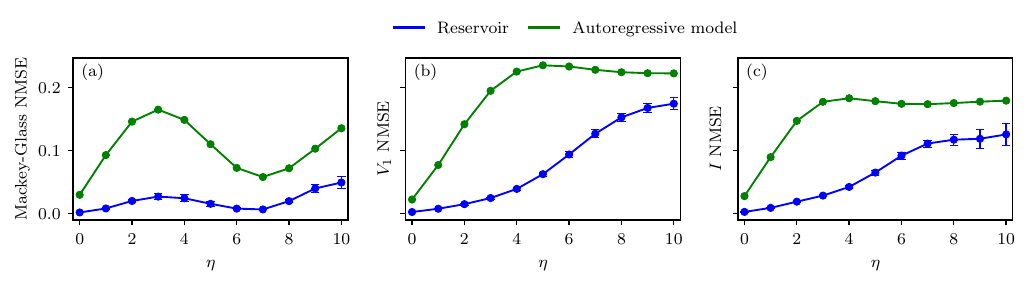}
\caption{Normalized mean squared error (NMSE) as a function of the prediction horizon $\eta$ for the (a) Mackey–Glass time series, and for the (b) $V_1$ and (c) $I$ components of the double-scroll system. The autoregressive model follows the setup described in this section and is provided with the same delayed feedback as the reservoir. The reservoir hyperparameters are the same as those used in the main text: $d=1$, $M=10^3$, and $K=45$. The error bars refer to standard deviation from five independent reservoir realizations.}\label{fig:MG_longterm}
\end{figure}

In the main text, we investigated the ability of the reservoir to forecast chaotic time series, focusing on the Mackey–Glass~\cite{Mackey1977} and double-scroll systems~\cite{DS}. In this section, we provide the details of how these time series were generated.

\subsection{Mackey-Glass equation}
The Mackey--Glass time series is the solution of the following delayed differential equation
\begin{equation}
\frac{\d}{\d t}u(t)
=
0.2\,\frac{u(t-17)}{1+u(t-17)^{10}}
-0.1\,u(t),
\end{equation}
where the delay is $17$ to ensure a chaotic behavior. The input sequence was obtained by numerically integrating the equation with a sampling interval of $\tau_r=3$, and subsequently rescaling the resulting values to the range $[0,1]$.

\subsection{Double-scroll equations}

Regarding the double-scroll system, we obtain the two time series in the main text (associated with the components $I$ and $V_1$) by solving the following system of differential equations:
\begin{align}
\dot{V}_1 &= V_1/R_1 - \Delta V/R_2 - 2 I_r \sinh(\beta \Delta V), \nonumber \\ 
\dot{V}_2 &= \Delta V/R_2 - 2 I_r \sinh(\beta \Delta V) - I, \nonumber \\ 
\dot{I} &= V_2 - R_4 I,
\end{align}
where $\Delta V=V_1-V_2$. Following the standard choice in the literature, we used the parameter values $(R_1, R_2, R_4)=(1.2, 3.44, 0.193)$, $\beta = 11.6$, and $I_r = 2.25 \times 10^{-5}$. Moreover, the system was integrated using a time step of $\Delta t=1$.

\subsection{Long-time forecasting}

In the main text, we show that the proposed reservoir can successfully perform the one-step-ahead prediction of the studied time series. We now show that this predictive capability is preserved as the forecasting horizon increases. Specifically, we consider the task of predicting the value of the time series $\eta$ time steps into the future, with the target defined as
\begin{equation}
\hat{o}_k = u_{k+\eta},
\end{equation}
where $\eta$ denotes the prediction horizon.

To quantify the accuracy, we use the normalized mean squared error (NMSE), defined as
\begin{equation}
\mathrm{NMSE} =
\frac{\sum_k \left(o_k-\hat{o}_k\right)^2}
{\sum_k \left(\hat{o}_k-\mu\right)^2},
\end{equation}
where $o_k$ is the predicted output, and $\mu$ is the mean of the target time series. 

To assess whether the proposed reservoir performs genuine nonlinear information processing, we compare its performance with that of a linear autoregressive model~\cite{box1994time}. Since the quantum reservoir employs a delayed feedback with a delay $d=1$, the autoregressive model is provided with the same temporal information. Its state is therefore defined as
\begin{equation}
\textbf{AR}_k = (u_k, u_{k-1}),
\end{equation}
which can be interpreted as a linear reservoir. The readout follows the same training and testing protocol adopted for the quantum reservoir, as described in the Methods section of the main text.

Figure~\ref{fig:MG_longterm} reports the NMSE as a function of the prediction horizon $\eta$ for all of the considered time series. The quantum reservoir consistently achieves a lower prediction error, whereas the autoregressive model rapidly degrades as $\eta$ increases.

\end{document}